\documentclass[prb,aps,twocolumn,showpacs,floats]{revtex4}
\usepackage{epsfig,bm,epsf,float,graphics}
\begin{document}
\draft
\title{EFFECTS OF FRUSTRATED SURFACE IN HEISENBERG THIN FILMS}
\author{V. Thanh Ngo$^{a,b}$ and H. T. Diep$^{a}$\footnote{ Corresponding author, E-mail:
diep@u-cergy.fr }}
\address{$^a$ Laboratoire de Physique Th\'eorique et Mod\'elisation,
CNRS-Universit\'e de Cergy-Pontoise, UMR 8089\\
2, Avenue Adolphe Chauvin, 95302 Cergy-Pontoise Cedex, France\\
$^b$ Institute of Physics, P.O. Box 429   Bo Ho, Hanoi 10000,
Vietnam}

\begin{abstract}
We study by extensive Monte Carlo (MC) simulations and analytical
Green function (GF) method effects of frustrated surfaces  on the
properties of thin films made of stacked triangular layers of
atoms bearing Heisenberg spins with an Ising-like interaction
anisotropy. We suppose that the in-plane surface interaction $J_s$
 can be antiferromagnetic  or ferromagnetic while all other interactions are
ferromagnetic. We show that the ground-state spin configuration is
non linear when $J_s$ is lower than a critical value $J_s^c$. The
film surfaces are then frustrated. In the frustrated case, there
are two phase transitions related to disorderings of surface and
interior layers. There is a good agreement between MC and GF
results. In addition, we show from MC histogram calculation that
the value of the ratio of critical exponents $\gamma/\nu$ of the
observed transitions is deviated from the values of two and three
Ising universality classes.  The origin of this deviation is
discussed with general physical arguments.

\end{abstract}
\pacs{} \maketitle
\section{Introduction}
This paper deals with the effect of the frustration in magnetic
thin films. The frustration is known to cause spectacular effects
in various bulk spin systems.  Its effects have been extensively
studied during the last decade theoretically, experimentally and
numerically. Frustrated model systems serve not only as
 testing ground for theories and approximations, but also to compare
with experiments.\cite{Diep2005}

On the other hand, surface physics and systems of nanoscales have
been also enormously studied during the last twenty years. This is
due in particular to applications in magnetic recording, let alone
fundamental theoretical interests.  Much is understood
theoretically and experimentally in  thin films where surfaces are
'clean' i.e. no impurities, no steps etc.
\cite{Binder-surf,Diehl,diep79,diep81,diep91,ngo2004}  Less is
known at least theoretically on complicated thin films with
special surface conditions such as defects\cite{Chung,Shen},
arrays of dots and magnetization reversal
phenomenon.\cite{Rousseau,Kong,Grolier,Jamet,Meyer,santamaria,santa1}

In this paper we study the frustration effect on properties of
thin films made of stacked triangular lattices.  In-plane
interaction of the surfaces is antiferromagnetic and that of
interior layers is ferromagnetic. The film surfaces are
frustrated.

The paper is organized as follows. Section II is devoted to the
description of our model.  The ground state in the case of
classical spins is determined as a function of the surface
interaction. In section III, we consider the case of quantum spins
and we apply the Green function technique to determine the layer
magnetizations and the transition temperature as a function of the
surface interaction. The classical ground state determined in
section II is used here as starting (input)configuration for
quantum spins. We are interested here in the effect of magnetic
frustration on magnetic properties of thin films. A phase diagram
is established showing interesting surface behaviors. Results from
Monte Carlo simulations for classical spins are shown in section
IV and compared to those obtained by the Green function method. We
also calculate by Monte Carlo histogram technique the critical
behavior of the phase transition observed here.   Concluding
remarks are given in Section V.

\section{Model}
It is known that many well-established theories failed to deal
with frustrated spin systems.\cite{Diep2005}  Among known striking
effects due to frustration, let us mention the high ground-state
(GS) degeneracy associated often with new symmetries which give
rise sometimes to new kinds of phase transition. One of the
systems which are most studied is the antiferromagnetic triangular
lattice. Due to its geometry, the spins are frustrated under
nearest-neighbor (NN) antiferromagnetic interaction. In the case
of Heisenberg model, the frustration results in a $120^\circ$ GS
structure: the NN spins form a $120^\circ$ angle  alternately in
the clockwise and counter-clockwise senses which are called left
and right chiralities.

\subsection{Hamiltonian}
In this paper we consider a thin film made up by stacking $N_z$
planes of triangular lattice of $L\times L$ lattice sites.

The Hamiltonian is given by
\begin{equation}
\mathcal H=-\sum_{\left<i,j\right>}J_{i,j}\mathbf S_i\cdot\mathbf
S_j -\sum_{<i,j>} I_{i,j}S_i^z S_j^z  \label{eqn:hamil1}
\end{equation}
where $\mathbf S_i$ is the Heisenberg spin at the lattice site
$i$, $\sum_{\left<i,j\right>}$ indicates the sum over the NN spin
pairs  $\mathbf S_i$ and $\mathbf S_j$.  The last term, which will
be taken to be very small,  is needed to make the film with a
finite thickness to have a phase transition at a finite
temperature in the case where all exchange interactions $J_{i,j}$
are ferromagnetic. This guarantees the existence of a phase
transition at finite temperature, since it is known that a
strictly two-dimensional system with an isotropic non-Ising spin
model (XY or Heisenberg model) does not have long-range ordering
at finite temperature.\cite{Mermin}

Interaction between two NN surface spins is equal to $J_s$.
Interaction between layers and interaction between NN in interior
layers are supposed to be ferromagnetic and all equal to $J=1$ for
simplicity. The two surfaces of the film are frustrated if $J_s$
is antiferromagnetic ($J_s<0$).
\subsection{Ground state}
In this paragraph, we suppose that the spins are classical.  The
classical ground state (GS) can be easily determined as shown
below.  Note that for antiferromagnetic systems, even for bulk
materials, the quantum GS cannot be exactly determined.  The
classical Neel state is often used as starting configuration for
quantum spins. We will follow the same line hereafter.

 For $J_s>0$
(ferromagnetic interaction), the magnetic GS is ferromagnetic.
However, when $J_s$ is negative the surface spins are frustrated.
Therefore, there is a competition between the non collinear
surface ordering and the ferromagnetic ordering due to the
ferromagnetic interaction from the spins of the beneath layer.

\begin{figure}[htb!]
\centerline{\epsfig{file=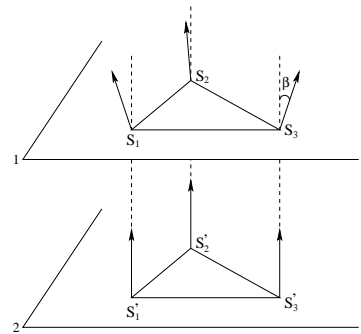,width=1.85in}} \caption{Non
collinear surface spin configuration. Angles between spins on
layer $1$ are all equal (noted $\alpha$), while angles between
vertical spins are $\beta$.} \label{fig:gsstruct}
\end{figure}

We first determine the GS configuration for $I=I_s=0.1$ by using
the steepest descent method : starting from a random spin
configuration, we calculate the magnetic local field at each site
and align the spin of the site in its local field. In doing so for
all spins and repeat until the convergence is reached, we obtain
in general the GS configuration, without metastable states in the
present model. The result shows that when $J_{s}$ is smaller than
a critical value $J_{s}^c$ the magnetic GS is obtained from the
planar $120^\circ$ spin structure, supposed to be in the $XY$
plane, by pulling them out of the spin $XY$ plane by an angle
$\beta$. The three spins on a triangle on the surface form thus an
'umbrella' with an angle $\alpha$ between them and an angle
$\beta$ between a surface spin and its beneath neighbor (see Fig.
\ref{fig:gsstruct}). This non planar structure is due to the
interaction of the spins on the beneath layer, just like an
external applied field in the $z$ direction. Of course, when $J_s$
is larger than $J_s^c$ one has the collinear ferromagnetic GS as
expected: the frustration is not strong enough to resist the
ferromagnetic interaction from the beneath layer.

\begin{figure}[htb!]
\centerline{\epsfig{file=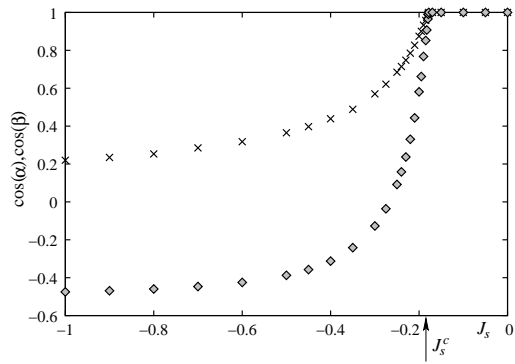,width=2.7in}}\caption{$\cos
(\alpha)$ (diamonds) and $\cos (\beta)$ (crosses) as functions of
$J_s$. Critical value of $J_s^c$ is shown by the arrow.}
\label{fig:gscos}
\end{figure}

We show in Fig. \ref{fig:gscos} $\cos(\alpha)$ and $\cos(\beta)$
as functions of $J_s$. The critical value $J_s^c$ is found between
-0.18 and -0.19.  This value can be calculated analytically by
assuming the 'umbrella structure'. For GS analysis, it suffices to
consider just a cell shown in Fig.\ref{fig:gsstruct}. This is
justified by the numerical determination discussed above.
Furthermore, we consider as a single solution all configurations
obtained from each other by any global spin rotation.

Let us consider the full Hamiltonian (\ref{eqn:hamil1}).  For
simplicity, the interaction inside the surface layer is set equal
$J_s$ $(-1 \leq J_s \leq 1)$ and all others are set equal to
$J>0$. Also, we suppose that $I_{i,j}=I_s$ for spins on the
surfaces with the same sign as $J_s$ and all other $I_{i,j}$ are
equal to $I>0$ for the inside spins including interaction between
a surface spin and the spin on the beneath layer.

The spins are numbered as in Fig. \ref{fig:gsstruct}: $S_1$, $S_2$
and $S_3$ are the spins in the surface layer (first layer),
$S'_1$, $S'_2$ and $S'_3$ are the spins in the internal layer
(second layer).  The Hamiltonian for the cell is written as
\begin{eqnarray}
H_p &=& -6\left[ J_s\left( \mathbf S_1\cdot \mathbf S_2 +\mathbf
S_2\cdot\mathbf S_3 + \mathbf S_3\cdot\mathbf S_1
\right)\right.\nonumber\\
&&+I_s\left( S^z_1S^z_2 + S^z_2S^z_3 + S^z_3S^z_1\right)\nonumber\\
&+&J\left(\mathbf S'_1\cdot \mathbf S'_2 +\mathbf
S'_2\cdot\mathbf S'_3 +\mathbf S'_3\cdot\mathbf S'_1\right)\nonumber\\
&&+I\left.\left( S'^z_1S'^z_2 + S'^z_2S'^z_3 + S'^z_3S'^z_1\right)\right] \nonumber \\
&-&2J\left( \mathbf S_1\cdot \mathbf S'_1 +\mathbf S_2\cdot\mathbf
S'_2 +\mathbf S_3\cdot\mathbf S'_3\right)\nonumber\\
&&-2I\left( S^z_1S'^z_1 + S'^z_2S'^z_2 + S^z_3S'^z_3\right),
\label{eqn:Hamilplaq}
\end{eqnarray}
Let us decompose each spin into two components: an $xy$ component,
which is a vector, and a $z$ component $\mathbf S_i=(\mathbf
S_i^{\parallel}, S_i^z)$. Only surface spins have $xy$ vector
components.  The angle between these $xy$ components of  NN
surface spins is $\gamma_{i,j}$ which is chosen by ($\gamma_{i,j}$
is in fact the projection of $\alpha $ defined above on the $xy$
plane)
\begin{equation}
\gamma_{1,2}=0,\ \gamma_{2,3}=\frac{2\pi}{3},\
\gamma_{3,1}=\frac{4\pi}{3}. \label{eqn:HSAngAlpha}
\end{equation}

The angles $\beta_i$ and $\beta'_i$ of the spin $\mathbf S_i$ and
$\mathbf S'_i$ with the $z$ axis are by symmetry
$$
\left\{%
\begin{array}{c}
\beta_1=\beta_2=\beta_3=\beta,\\
\beta'_1=\beta'_2=\beta'_3=0,\\
\end{array}
\right.
$$

The total energy of the cell (\ref{eqn:Hamilplaq}), with $S_i =
S'_i = \frac{1}{2}$, can be rewritten as
\begin{eqnarray}
H_p&=&-\frac{9(J+I)}{2} -\frac{3(J+I)}{2}\cos\beta
-\frac{9(J_s+I_s)}{2}\cos^2\beta
\nonumber\\
&+&\frac{9J_s}{4}\sin^2\beta. \label{eqn:totEplaq}
\end{eqnarray}
By a variational method,  the minimum of the cell energy
corresponds to
\begin{equation}
\frac{\partial H_p}{\partial\beta} =\left( \frac{27}{2}J_s+
9I_s\right)\cos\beta\sin\beta +\frac{3}{2}(J+I)\sin\beta \ = \ 0
\label{eqn:DerivE}
\end{equation}
We have
\begin{equation}
\cos\beta = -\frac{J+I}{9J_s+6I_s}. \label{eqn:GSsolu}
\end{equation}

For  given values of $I_s$ and $I$, we see that the solution
(\ref{eqn:GSsolu}) exists for $J_s \leq J_s^c$ where the critical
value $J_s^c$ is determined by $-1\leq \cos\beta \leq 1$. For
$I=-I_s=0.1$, $J_s^c \approx -0.1889 J $
in excellent agreement with the numerical result.

The classical GS determined here will be used as input GS
configuration for quantum spins considered in the next section.

\section{Green function method}

Let us consider the quantum spin case.  For a given value of
$J_s$, we shall use the Green function method to calculate the
layer magnetizations as functions of temperature. The details of
the method in the case of non collinear spin configuration have
been given in Ref.\cite{Rocco}. We briefly recall it here and show
the application to the present model.
\subsection{Formalism}
We can rewrite the full Hamiltonian (\ref{eqn:hamil1}) in the
local framework of the classical GS configuration as
\begin{eqnarray}
\mathcal H &=& - \sum_{<i,j>}
J_{i,j}\Bigg\{\frac{1}{4}\left(\cos\theta_{ij} -1\right)
\left(S^+_iS^+_j +S^-_iS^-_j\right)\nonumber\\
&+& \frac{1}{4}\left(\cos\theta_{ij} +1\right) \left(S^+_iS^-_j
+S^-_iS^+_j\right)\nonumber\\
&+&\frac{1}{2}\sin\theta_{ij}\left(S^+_i +S^-_i\right)S^z_j
-\frac{1}{2}\sin\theta_{ij}S^z_i\left(S^+_j
+S^-_j\right)\nonumber\\
&+&\cos\theta_{ij}S^z_iS^z_j\Bigg\}- \sum_{<i,j>}I_{i,j}S^z_iS^z_j
\label{eq:HGH2}
\end{eqnarray}
where $\cos\left(\theta_{ij}\right)$ is the angle between two NN
spins determined classically in the previous section.

Following Tahir-Kheli and ter Haar,\cite{tahir} we define two
double-time Green functions by
\begin{eqnarray}
G_{ij}(t,t')&=& \ll S^{+}_i(t) ; S^-_j(t')\gg, \\
F_{ij}(t,t')&=&\ll S^{-}_i(t) ; S^+_j(t')\gg.
\end{eqnarray}

The equations of motion for $G_{ij}(t,t')$ and $F_{ij}(t,t')$ read
\begin{eqnarray}
i\frac {d}{dt}G_{i,j}\left( t,t'\right) &=& \left<\left[ S^+_i
\left( t\right) , S^-_j \left( t'\right)\right]\right>\delta\left(
t-t'\right) \nonumber\\
&-& \left<\left< \left[\mathcal H, S^+_i\left( t\right)\right] ;
S^-_j \left( t'\right) \right>\right>,
\label{eq:HGEoMG}\\
i\frac {d}{dt}F_{i,j}\left( t,t'\right) &=& \left<\left[ S^-_i
\left( t\right) , S^-_j \left( t'\right)\right]\right>\delta\left(
t-t'\right)\nonumber \\
&-& \left<\left< \left[\mathcal H, S^-_i\left( t\right)\right] ;
S^-_j \left( t'\right) \right>\right>, \label{eq:HGEoMF}
\end{eqnarray}

We neglect higher order correlations by using the Tyablikov
decoupling scheme\cite{Tyablikov} which is known to be valid for
exchange terms.\cite{fro}   Then, we introduce the Fourier
transforms

\begin{eqnarray}
G_{i, j}\left( t, t'\right) &=& \frac {1}{\Delta}\int\int d\mathbf
k_{xy}\frac{1}{2\pi}\int^{+\infty}_{-\infty}d\omega e^{-i\omega
\left(t-t'\right)}.\nonumber\\
&&\hspace{0.7cm}g_{n,n'}\left(\omega , \mathbf k_{xy}\right)
e^{i\mathbf k_{xy}\cdot \left(\mathbf R_i-\mathbf
R_j\right)},\label{eq:HGFourG}\\
F_{i, j}\left( t, t'\right) &=& \frac {1}{\Delta}\int\int d\mathbf
k_{xy}\frac{1}{2\pi}\int^{+\infty}_{-\infty}d\omega e^{-i\omega
\left(t-t'\right)}.\nonumber\\
&&\hspace{0.7cm}f_{n,n'}\left(\omega , \mathbf k_{xy}\right)
e^{i\mathbf k_{xy}\cdot \left(\mathbf R_i-\mathbf
R_j\right)},\label{eq:HGFourF}
\end{eqnarray}
where $\omega$ is the spin-wave frequency, $\mathbf k_{xy}$
denotes the wave-vector parallel to $xy$ planes, $\mathbf R_i$ is
the position of the spin at the site $i$, $n$ and $n'$ are
respectively the index of the layers where the sites $i$ and $j$
belong to. The integral over $\mathbf k_{xy}$ is performed in the
first Brillouin zone whose surface is $\Delta$ in the $xy$
reciprocal plane.

The Fourier transforms of the retarded Green functions satisfy a
set of equations rewritten under a matrix form
\begin{equation}
\mathbf M \left( \omega \right) \mathbf g = \mathbf u,
\label{eq:HGMatrix}
\end{equation}
where $\mathbf M\left(\omega\right)$ is a square matrix
$\left(2N_z \times 2N_z\right)$, $\mathbf g$ and $\mathbf u$ are
the column matrices which are defined as follows
\begin{equation}
\mathbf g = \left(%
\begin{array}{c}
  g_{1,n'} \\
  f_{1,n'} \\
  \vdots \\
  g_{N_z,n'} \\
  f_{N_z,n'} \\
\end{array}%
\right) ,\hspace{1truecm} \mathbf u =\left(%
\begin{array}{c}
  2 \left< S^z_1\right>\delta_{1,n'} \\
  0 \\
  \vdots \\
  2 \left< S^z_{N_z}\right>\delta_{N_z,n'} \\
  0 \\
\end{array}%
\right) , \label{eq:HGMatrixgu}
\end{equation}
and
\begin{equation}
\mathbf M\left(\omega\right) = \left(%
\begin{array}{ccccc}
  A^+_1    & B_1    &  D^+_1 &  D^-_1 & \cdots \\
  -B_1     & A^-_1  & -D^-_1 & -D^+_1 & \vdots \\
   \vdots  & \cdots & \cdots & \cdots &\vdots\\
  \vdots   & C^+_{N_z}   & C^-_{N_z}   & A^+_{N_z}      & B_{N_z}\\
  \cdots        & -C^-_{N_z}  & -C^+_{N_z}  & -B_{N_z}       & A^-_{N_z}\\
\end{array}%
\right), \label{eq:HGMatrixM}
\end{equation}
where
\begin{eqnarray}
A_n^\pm &=&  \omega \pm\Big[\frac{1}{2}J_n \left< S^z_n\right>
\left(Z\gamma\right)\left(\cos\theta_{n} +1\right)\nonumber\\
&-& J_n \left< S^z_n\right>Z\cos\theta_{n} -J_{n, n+1}\left< S^z_{n+1}\right>\cos\theta_{n,n+1} \nonumber\\
&-& J_{n, n-1}\left< S^z_{n-1}\right>\cos\theta_{n,n-1} -Z
I_{n} \left< S^z_n\right>\nonumber\\
&&\ -\ I_{n,n+1}\left< S^z_{n+1}\right>-I_{n,n-1} \left<
S^z_{n-1}\right>\Big],\\
B_n &=& \frac{1}{2}J_{n}\left<
S^z_n\right>\left(\cos\theta_{n}-1\right)\left(Z\gamma\right),\\
C_n^\pm &=& \frac{1}{2}J_{n,n-1}\left<
S^z_n\right>\left(\cos\theta_{n,n-1}\pm 1\right),\\
D_n^\pm &=& \frac{1}{2}J_{n,n+1}\left<
S^z_n\right>\left(\cos\theta_{n,n+1}\pm 1\right),
\end{eqnarray}
in which, $Z=6$ is the number of in-plane NN, $\theta_{n,n\pm 1}$
the angle between two NN spins belonging to the layers $n$ and
$n\pm1$, $\theta_{n}$ the angle between two in-plane NN in the
layer $n$, and
$$\gamma =\left[ 2\cos \left( k_x a \right)
+ 4\cos \left( k_y a/2 \right)\cos\left( k_y
a\sqrt{3}/2\right)\right]/Z.$$  Here, for compactness we have used
the following notations:

i) $J_n$ and $I_n$ are the in-plane interactions. In the present
model $J_n$ is equal to $J_s$ for the two surface layers and equal
to $J$ for the interior layers. All $I_n$ are set to be $I$.

ii) $ J_{n,n\pm 1}$ and $ I_{n,n\pm 1}$ are the interactions
between a spin in the $n^{th}$ layer and its neighbor in the
$(n\pm 1)^{th}$ layer. Of course, $ J_{n,n-1}=I_{n,n-1}$=0 if
$n=1$, $ J_{n,n+1}=I_{n,n+1}$=0 if $n=N_z$.

Solving det$|\mathbf M|=0$, we obtain the spin-wave spectrum
$\omega$ of the present system.  The solution for the Green
function $g_{n,n}$ is given by
\begin{equation}
g_{n,n} = \frac{\left|\mathbf M\right|_n}{\left|\mathbf M\right|},
\end{equation}
where $\left|\mathbf M\right|_n$ is the determinant made by
replacing the $n$-th column of $\left|\mathbf M\right|$ by
$\mathbf u$ in (\ref{eq:HGMatrixgu}). Writing now
\begin{equation}
\left|\mathbf M\right| = \prod_i \left(\omega -
\omega_i\left(\mathbf k_{xy}\right)\right),
\end{equation}
one sees that $\omega_i\left(\mathbf k_{xy}\right) ,\ i = 1,\cdots
,\ N_z$, are poles of the Green function $g_{n,n}$.
$\omega_i\left(\mathbf k_{xy}\right)$ can be obtained by solving
$\left|\mathbf M\right|=0$. In this case, $g_{n,n}$ can be
expressed as
\begin{equation}
g_{n, n} = \sum_i\frac {h_n\left(\omega_i\left(\mathbf
k_{xy}\right)\right)}{\left( \omega - \omega_i\left(\mathbf
k_{xy}\right)\right)}, \label{eq:HGGnn}
\end{equation}
where $h_n\left(\omega_i\left(\mathbf k_{xy}\right)\right)$ is
\begin{equation}
h_n\left(\omega_i\left(\mathbf k_{xy}\right)\right) = \frac{\left|
\mathbf M\right|_n \left(\omega_i\left(\mathbf
k_{xy}\right)\right)}{\prod_{j\neq i}\left(\omega_j\left(\mathbf
k_{xy}\right)-\omega_i\left(\mathbf k_{xy}\right)\right)}.
\end{equation}

Next, using the spectral theorem which relates the correlation
function \(\langle S^-_i S^+_j\rangle \) to the Green
functions,\cite{zu} one has
\begin{eqnarray}
\left< S^-_i S^+_j\right> &=& \lim_{\varepsilon\rightarrow 0}
\frac{1}{\Delta}\int\int d\mathbf k_{xy}
\int^{+\infty}_{-\infty}\frac{i}{2\pi}\big( g_{n, n'}\left(\omega
+ i\varepsilon\right)\nonumber\\
&-& g_{n, n'}\left(\omega - i\varepsilon\right)\big)
\frac{d\omega}{e^{\beta\omega} - 1}e^{i\mathbf
k_{xy}\cdot\left(\mathbf R_i -\mathbf R_j\right)},
\end{eqnarray}
where $\epsilon$ is an  infinitesimal positive constant and
$\beta=1/k_BT$, $k_B$ being the Boltzmann constant.  For spin
$S=1/2$, the thermal average of the $z$ component of the $i-th$
spin belonging to the $n-th$ layer is given by

\begin{equation}
\left< S^z_i\right> =\frac{1}{2}-\left< S^-_i S^+_i\right>
\end{equation}
In the following we shall use the case of spin one-half.  Note
that for the case of general $S$, the expression for $\left<
S^z_i\right>$ is more complicated since it involves higher
quantities such as $\left< (S^z_i)^2\right>$.

Using the Green function presented above, we can calculate
self-consistently various physical quantities as functions of
temperature $T$. The first important quantity is the temperature
dependence of the angle between each spin pair.  This can be
calculated in a self-consistent manner at any temperature by
minimizing the free energy at each temperature to get the correct
value of the angle as it has been done for a frustrated bulk spin
systems.\cite{santa2}  In this paper, we limit ourselves to the
self-consistent calculation of the layer magnetizations which
allows us to establish the phase diagram as seen in the following.

For numerical calculation, we used $I=0.1J$ with $J=1$.  For
positive $J_s$, we take $I_s=0.1$ and for negative $J_s$, we use
$I_s=-0.1$.  A size of $80^2$ points in the first Brillouin zone
is used for numerical integration. We start the self-consistent
calculation from $T=0$ with a small step for temperature $5\times
10^{-3}$ or $10^{-1}$ (in units of $J/k_B$). The convergence
precision has been fixed at the fourth figure of the values
obtained for the layer magnetizations.

\subsection{Phase transition and phase diagram of the quantum case}

\begin{figure}[hbt!]
\centerline{\epsfig{file=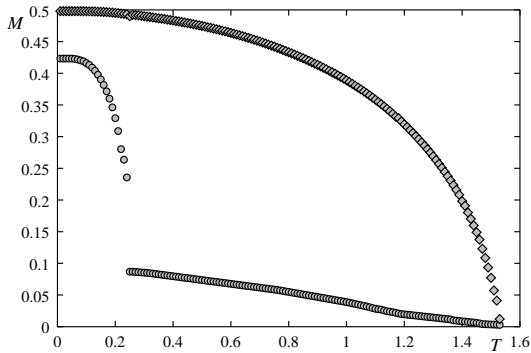,width=2.8in}} \caption{First
two layer-magnetizations obtained by the Green function technique
vs. $T$ for $J_{s} = -0.5$ with $I=-I_s=0.1$. The surface-layer
magnetization (lower curve) is much smaller than the second-layer
one. See text for comments.} \label{fig:HGn05Ms}
\end{figure}

We first show an example where $J_{s} = -0.5$ in Fig.
\ref{fig:HGn05Ms}. As seen, the surface-layer  magnetization is
much smaller than the second-layer one. In addition there is a
strong spin contraction at $T=0$ for the surface layer. This is
due to the antiferromagnetic nature of the in-plane surface
interaction $J_s$.  One sees that the surface becomes disordered
at a temperature $T_1\simeq 0.2557$ while the second layer remains
ordered up to $T_2\simeq 1.522$.   Therefore, the system is
partially disordered for temperatures between $T_1$ and $T_2$.
 This result is very interesting because it confirms again the
existence of the partial disorder in quantum spin systems observed
earlier in bulk frustrated quantum spin
systems.\cite{Rocco,santa2}  Note that between $T_1$ and $T_2$,
the ordering of the second layer acts as an external field on the
first layer, inducing therefore a small value of its
magnetization.  A further evidence of the existence of the surface
transition will be provided with the surface susceptibility in the
MC results shown below.

\begin{figure}[hbt!]
\centerline{\epsfig{file=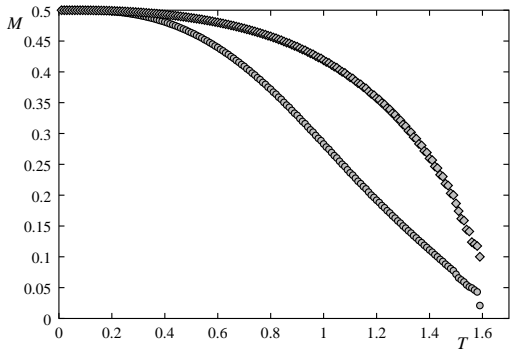,width=2.7in}} \caption{First
two layer-magnetizations obtained by the Green function technique
vs. $T$ for $J_{s} = 0.5$ with $I=I_s=0.1$.} \label{fig:HGp05Ms}
\end{figure}

Figure \ref{fig:HGp05Ms} shows the non frustrated case where
$J_s=0.5$, with $I=I_s=0.1$.  As seen, the first-layer
magnetization is smaller than the second-layer one. There is only
one transition temperature. Note the difficulty for numerical
convergency when the magnetizations come close to zero.

\begin{figure}[hbt!]
\centerline{\epsfig{file=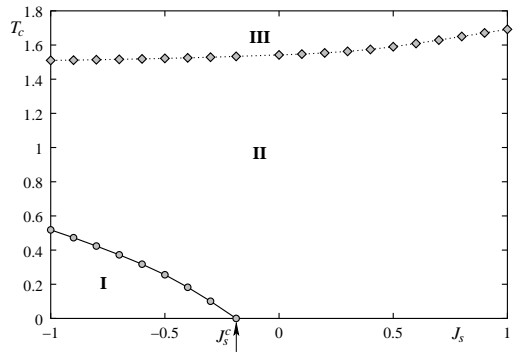,width=2.7in}} \caption{Phase
diagram in the space ($J_{s},T$) for the quantum Heisenberg model
with $N_z=4$, $I=|I_s|=0.1$. See text for the description of
phases I to III.} \label{fig:HGDG}
\end{figure}

We show in Fig. \ref{fig:HGDG} the phase diagram in the space
$(J_s,T)$.  Phase I denotes the ordered phase with surface non
collinear spin configuration, phase II indicates the collinear
ordered state, and phase III is the paramagnetic phase. Note that
the surface transition does not exist for $J_s \geq J_s^c$.

\section{Monte Carlo results}

It is known that methods for quantum spins, such as spin-wave
theory or Green function method presented above, suffer at high
temperatures. Spin-wave theory, even with magnon-magnon
interactions taken into account, cannot go to temperatures close
to $T_c$. Green function method on the other hand can go up to
$T_c$ but due to the decoupling scheme, it cannot give correct
critical behavior at $T_c$.  Fortunately, we know that quantum
spin systems behave as their classical counterparts at high $T$.
So, to see if the phase diagram obtained in the previous section
for the quantum model is precise or not,  we can consider its
classical version and use MC simulations to obtain the phase
diagram for comparison.  MC simulations are excellent means to
overcome approximations used in analytic calculations for the high
$T$ region  as discussed above.

In this paragraph, we show the results obtained by MC simulations
with the Hamiltonian (\ref{eqn:hamil1}) but the spins are the
classical Heisenberg model of magnitude $S=1$.

 The film sizes are $L\times L \times N_z$ where
$N_z=4$ is the number of layers (film thickness) taken as in the
quantum case presented above. We use here $L=24, 36, 48, 60$ to
study finite-size effects as shown below. Periodic boundary
conditions are used in the $XY$ planes. The equilibrating time is
about $10^6$ MC steps per spin and the averaging time is $2\times
10^6$ MC steps per spin. $J=1$ is taken as unit of energy in the
following.

\begin{figure}[thb!]
\centerline{\epsfig{file=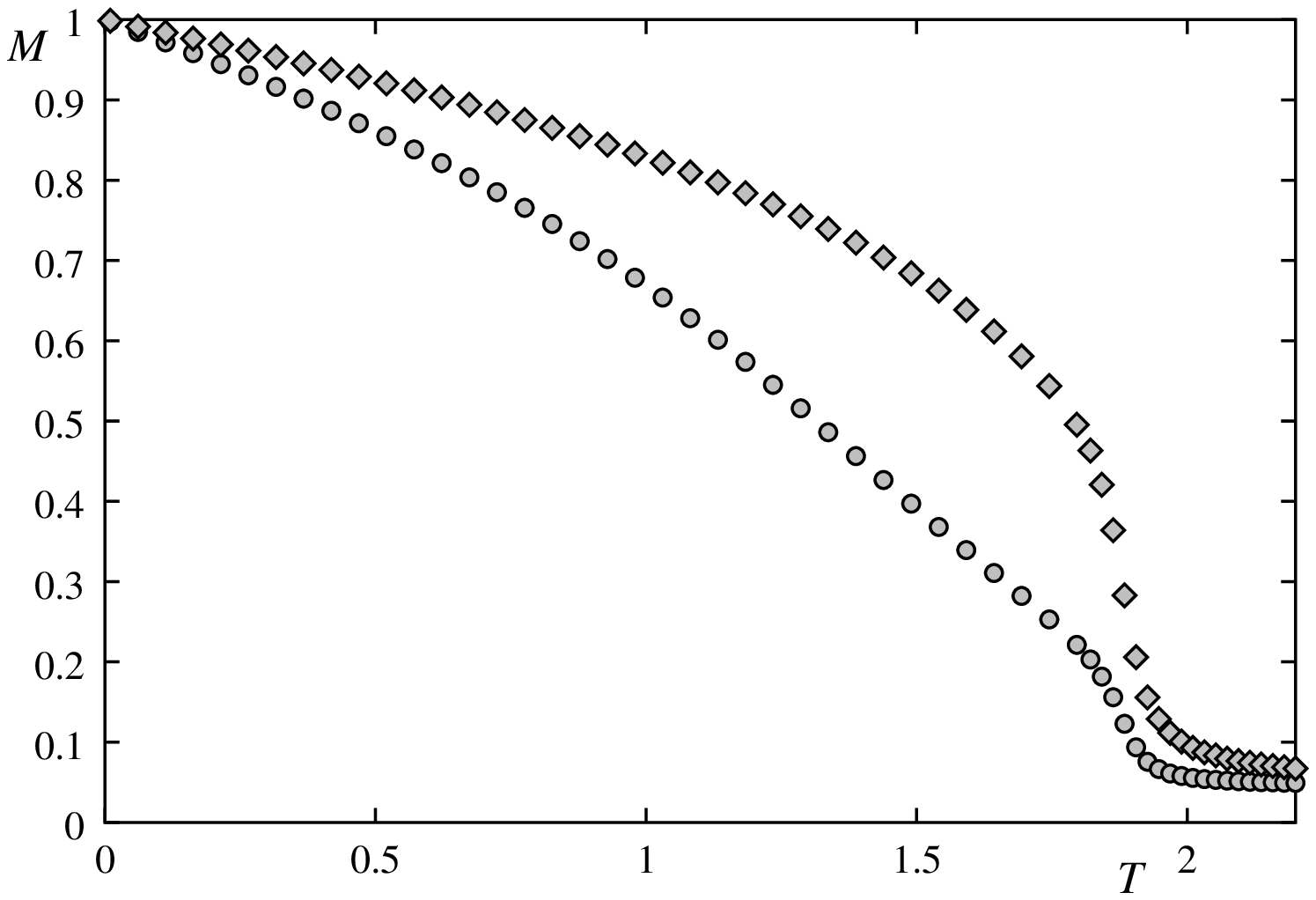,width=2.7in}}
\caption{Magnetizations of layer 1 (circles) and layer 2
(diamonds) versus temperature $T$ in unit of $J/k_B$ for $J_s=0.5$
with $I=I_s=0.1$, $L=36$.} \label{fig:HSp05Ms}
\end{figure}

Let us show in Fig. \ref{fig:HSp05Ms} the layer magnetization of
the first two layers as a function of $T$ , in the case $J_s=0.5$
with $N_z=4$ (the third and fourth layers are symmetric). Though
we observe a smaller magnetization for the surface layer, there is
clearly no surface transition just as in the quantum case.

\begin{figure}[thb!]
\centerline{\epsfig{file=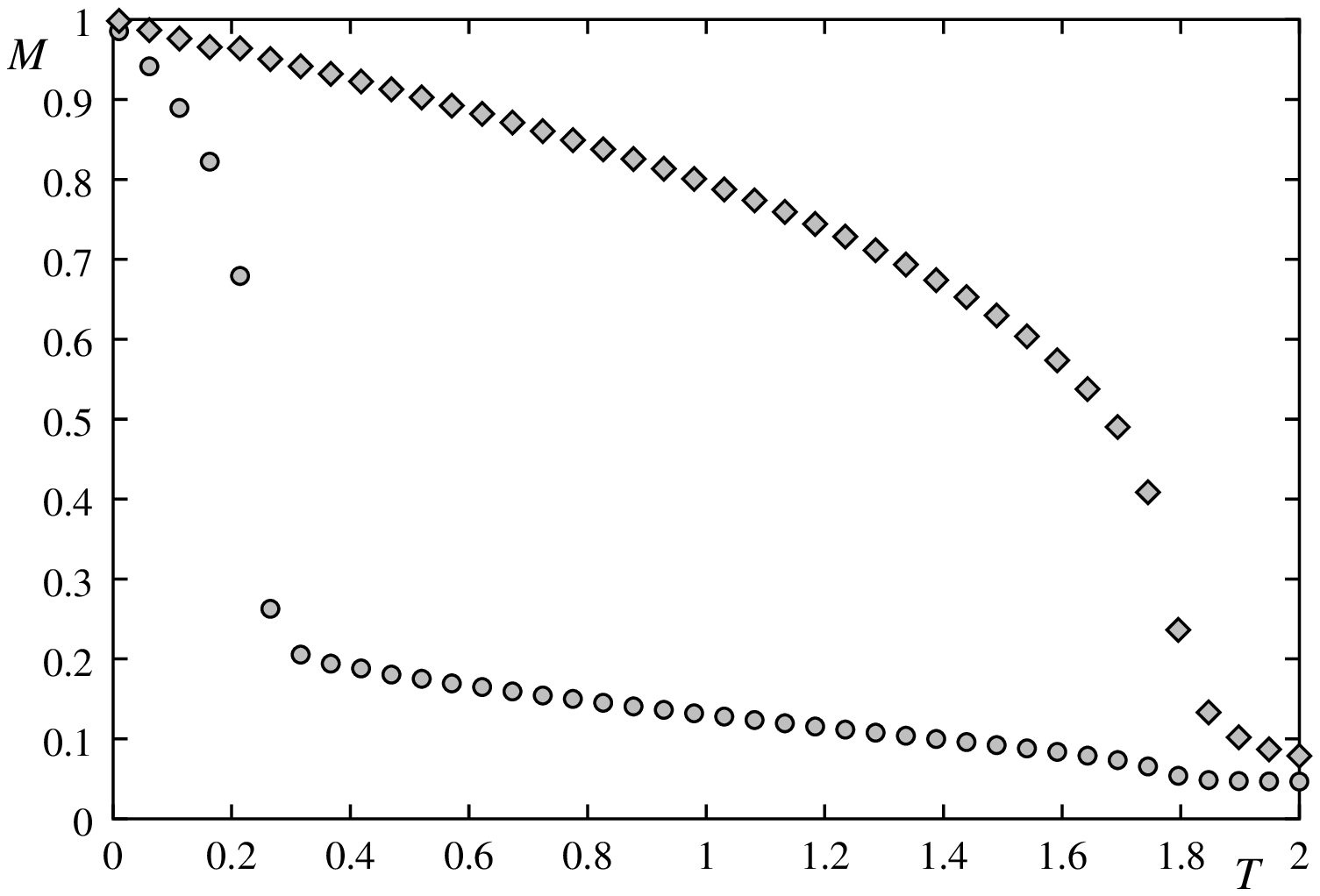,width=2.7in}}
\caption{Magnetizations of layer 1 (circles) and layer 2
(diamonds) versus temperature $T$ in unit of $J/k_B$ for
$J_s=-0.5$ with $I=-I_s=0.1$, $L=36$.} \label{fig:HSn05Ms}
\end{figure}

In Fig. \ref{fig:HSn05Ms} we show a frustrated case where
$J_s=-0.5$.  The surface layer in this case becomes disordered at
a temperature much lower than that for the second layer.  Note
that the surface magnetization is not saturated to 1 at $T=0$.
This is  because the surface spins make an angle with the $z$ axis
so their $z$ component is less than 1 in the GS.

\begin{figure}[hbt!]
\centerline{\epsfig{file=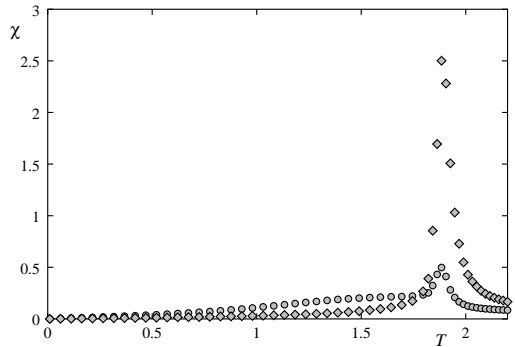,width=2.7in}}
\caption{Susceptibilities of layer 1 (circles) and layer 2
(diamonds) versus temperature $T$ in unit of $J/k_B$ for $J_s=0.5$
with $I=I_s=0.1$, $L=36$.} \label{fig:HSp05Ss}
\end{figure}

\begin{figure}[hbt!]
\centerline{\epsfig{file=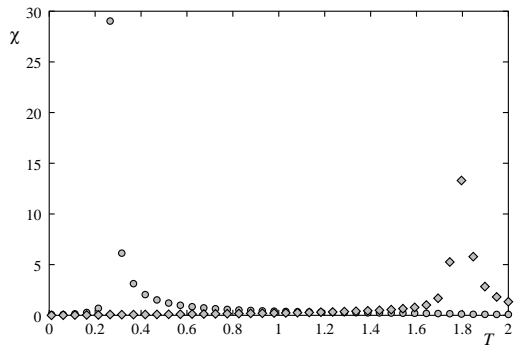,width=2.7in}}
\caption{Susceptibility of layer 1 (circles) and layer 2
(diamonds) versus temperature $T$ in unit of $J/k_B$ for
$J_s=-0.5$ with $I=-I_s=0.1$, $L=36$. Note that for clarity, the
susceptibility of the layer 2 has been multiplied by a factor 5.}
\label{fig:HSn05Ss}
\end{figure}

Figures \ref{fig:HSp05Ss}  shows the susceptibilities of the first
and second layers in the case where $J_s=0.5$ with $I=I_s=0.1$
where one observes the peaks at the same temperature indicating a
single transition in contrast to the frustrated case shown in Fig.
\ref{fig:HSn05Ss}. These results confirm the above results of
layer magnetizations.


\begin{figure}[hbt!]
\centerline{\epsfig{file=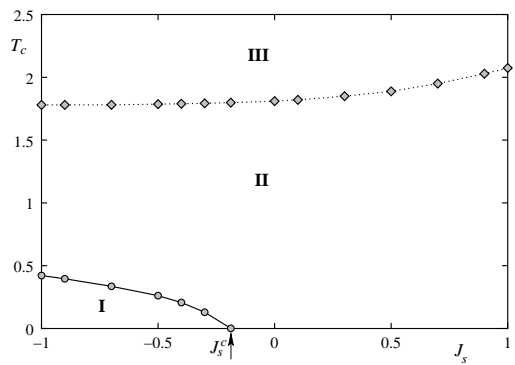,width=2.7in}} \caption{Phase
diagram in the space ($J_s,T$) for the classical Heisenberg model
with $N_z=4$, $I=|I_s|=0.1$. Phases I to III have the same
meanings as those in Fig. \ref{fig:HGDG} .} \label{fig:HSDG}
\end{figure}

To establish the phase diagram, the transition temperatures are
taken at the change of curvature of the layer magnetizations, i.e.
at the maxima of layer susceptibilities shown before. Figure
\ref{fig:HSDG} shows the phase diagram obtained in the space
$(J_s,T)$.  Interesting enough, this phase diagram resembles
remarkably to that obtained for the quantum counterpart model
shown in Fig. \ref{fig:HGDG}.

Let us study the finite size effect of the phase transitions shown
in Fig. \ref{fig:HSDG}.  To this end we use the histogram
technique which has been proved so far to be excellent for the
calculation of critical
exponents.\cite{Ferrenberg1,Ferrenberg2,Ferrenberg3} The principle
is as follows. Using the Metropolis algorithm to determine
approximately the critical temperature region, then choosing a
temperature $T_0$ as close as possible to the presupposed
transition temperature. We then make a very long run at $T_0$ to
establish an energy histogram.  From formulae established using
the statistical canonical distribution, we can calculate physical
quantities in a continuous manner for temperatures around $T_0$.
\cite{Ferrenberg1,Ferrenberg2,Ferrenberg3}  We do not have problem
to identify the transition temperature as well as the maximal
values of fluctuation quantities such as specific heat and
susceptibility.

\begin{figure}[hbt!]
\centerline{\epsfig{file=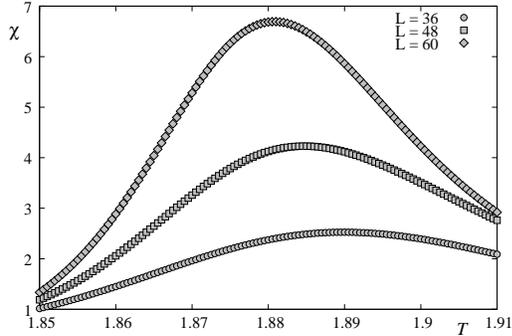,width=2.7in}}
 \caption{Susceptibility versus $T$ for $L=36,48,60$ with
$J_s=0.5$ and $I=I_s=0.1$.} \label{fig:sus3p05sl}
\end{figure}

Figure \ref{fig:sus3p05sl} shows the susceptibility versus $T$ for
$L=36, 48, 60$ in the case of $J_s=0.5$. For presentation
convenience, the size $L=24$ has been removed since the peak for
this case is rather flat in the scale of the figure. However, it
shall be used to calculate the critical exponent $\gamma$ for the
transition. As seen, the maximum $\chi^{\max}$ of the
susceptibility increases with increasing $L$.

\begin{figure}[hbt!]
\centerline{\epsfig{file=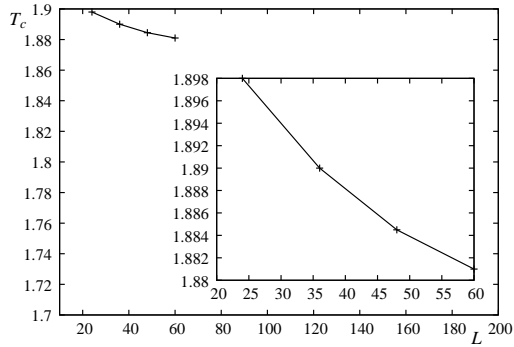,width=2.7in}}
\caption{Transition temperature versus $L$ with $J_s=0.5$ and
$I=I_s=0.1$.  The inset shows the enlarged scale.}
\label{fig:stcp05}
\end{figure}

For completeness, we show in Fig. \ref{fig:stcp05} the transition
temperature as a function of $L$.  A rough extrapolation to
infinity gives $T_c^{\infty}\simeq 1.86\pm 0.02$.

\begin{figure}[hbt!]
\centerline{\epsfig{file=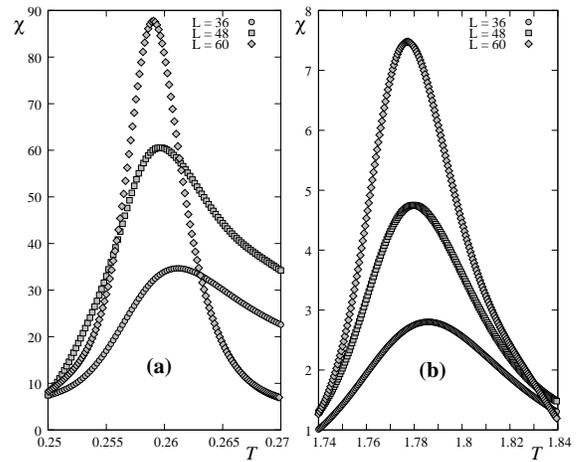,width=3.in}}
 \caption{Layer susceptibilities versus $T$ for $L=36,48,60$ with
$J_s=-0.5$ and $I=-I_s=0.1$. Left (right) figure corresponds to
the first (second) layer susceptibility.} \label{fig:sus3n05sl}
\end{figure}

In the frustrated case, i.e.  $J_s < J_s^c$, we perform the same
calculation for finite-size effect.  Note that in this case there
are two phase transitions.  We show in Fig. \ref{fig:sus3n05sl}
the layer susceptibilities as functions of $T$ for different $L$.
As seen, both surface and second-layer transitions have a strong
size dependence.

\begin{figure}[hbt!]
\centerline{\epsfig{file=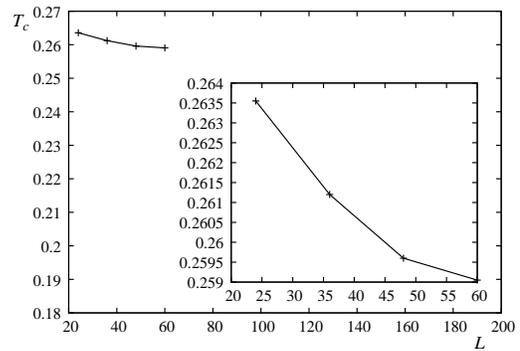,width=2.7in}}
\caption{Transition temperature versus $L$ for the surface layer
in the case $J_s=-0.5$ with $I=-I_s=0.1$. The inset shows the
enlarged scale.} \label{fig:stcn05_T1}
\end{figure}

\begin{figure}[hbt!]
\centerline{\epsfig{file=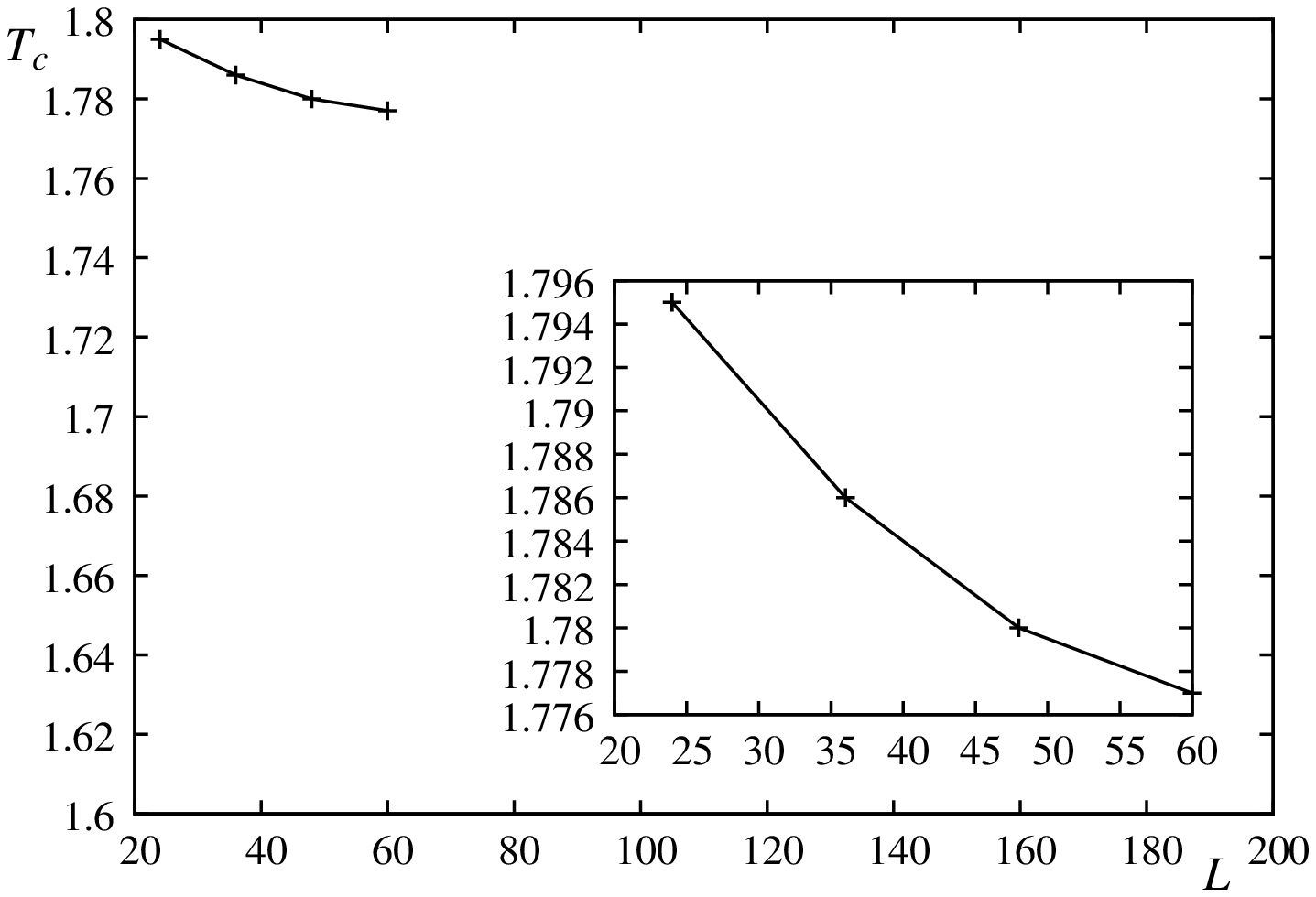,width=2.7in}}
\caption{Transition temperature versus $L$ for the second layer in
the case $J_s=-0.5$ with $I=-I_s=0.1$. The inset shows the
enlarged scale.} \label{fig:stcn05_T2}
\end{figure}

We show in Fig. \ref{fig:stcn05_T1} and \ref{fig:stcn05_T2} the
size dependence of the transition temperatures $T_1$ (surface
transition) and $T_2$ (second-layer transition).

\begin{figure}[thb!]
\centerline{\epsfig{file=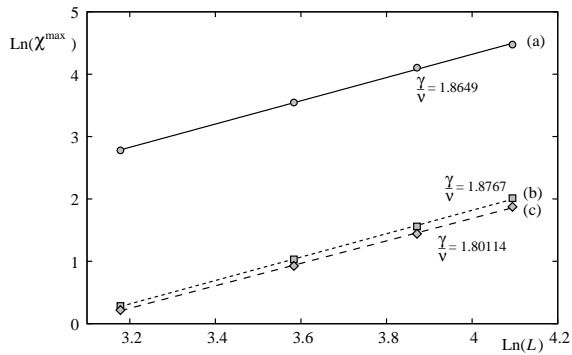,width=3.0in}} \caption{Maximum
of surface-layer susceptibility versus $L$ for $L=24,36,48,60$
with $J_s=-0.5$ (a,b), $J_s =0.5$ (c) and $I=|I_s|=0.1$, in the
$\ln - \ln$ scale. The slope gives $\gamma/\nu$ indicated in the
figure for each case. See text for comments.} \label{fig:gamma}
\end{figure}

The size dependence of the maxima observed above allows us to
estimate the ratio $\gamma/\nu$.  We show now $\ln \chi^{\max}$ as
a function of $\ln L$ for the different cases studied above.
Figure \ref{fig:gamma}(a) and (b) correspond respectively to the
transitions of surface and second layer occurring in the
frustrated case with $J_s=-0.5$, while Figure \ref{fig:gamma}(c)
corresponds to the unique transition occurring in the non
frustrated case with $J_s=0.5$.  The slopes of these straight
lines give $\gamma/\nu \simeq 1.864\pm 0.034$ (a), $1.878\pm
0.027$ (b), $ 1.801\pm 0.027$ (c). The errors were estimated from
the mean least-square fitting and errors on the peak values
obtained with different values of $T_0$ (multi histogram
technique).   Within errors, the first two values, which
correspond to the frustrated case, can be considered as identical,
while the last one corresponding to the non frustrated case is
different.  We will return to this point later.

At this stage, we would like to emphasize the following points.
First, we observe that these values of $\gamma/\nu$ are found to
be between those of the two-dimensional (2D) Ising model
($\gamma/\nu=1.75$) and the three-dimensional one
($\gamma/\nu=1.241/0.63 \simeq 1.97$). A question which naturally
arises on the role of the Ising-like anisotropy term, of the
two-fold chiral symmetry  and of the film thickness. The role of
the anisotropy term and the chiral symmetry is obvious: the Ising
character should be observed in the result (we return to this
point below). It is however not clear for the effect of the
thickness. Some arguments, such as those from renormalization
group, say that the correlation length in the direction
perpendicular  to the film is finite, hence it is irrelevant to
the criticality, the 2D character therefore should be
theoretically preserved. We think that such arguments are not
always true because it is difficult to conceive that when the film
thickness becomes larger and larger the 2D universality should
remain. Instead, we believe that that the finite thickness of the
film affects the 2D universality in one way or another, giving
rise to "effective" critical exponents with values deviated from
the 2D ones.  The larger the thickness is, the stronger this
deviation becomes.  The observed values of $\gamma/\nu$ shown
above may contain an effect of a 2D-3D cross-over. At this point,
we would like to emphasize that, in the case of simple surface
conditions, i.e. no significant deviation of the surface
parameters with respect to those of the bulk,  the bulk behavior
is observed when the thickness becomes larger than a dozen of
atomic layers:\cite{diep79,diep81} surface effects are
insignificant on thermodynamic properties of the film.  There are
therefore reasons to believe that there should be a cross-over
from 2D to 3D at some film thickness.  Of course, this is an
important issue which needs to be theoretically clarified in the
future.  We return now to the effect of Ising anisotropy and
chiral symmetry. The deviation from the 2D values may result in
part from a complex coupling between the Ising symmetry, due to
anisotropy and chirality,  and the continuous nature of the
classical Heisenberg spins studied here.  This deviation may be
important if the anisotropy constant $I$ is small.  For the effect
chiral symmetry, it is  a complex matter.  To show the complexity
in determining the critical universality with chiral symmetry, let
us discuss about a simpler case with similar chiral symmetry: the
XY model on the fully frustrated Villain's lattice. There has been
a large number of investigations on the nature of the transition
observed in this 2D case in the context of the frustration
effects.\cite{Diep2005,Olsson2005,Boubcheur1998} In this model,
though the chirality symmetry argument says that the transition
should be of 2D Ising universality class, many investigators found
a deviation of critical exponents from those of the 2D case (see
review in Ref. \cite{Boubcheur1998}). For example, the following
values are found for the critical exponent $\nu$:
$\nu=0.889$\cite{Ramirez94} and $\nu=0.813$.\cite{Lee94} These
values are close to that obtained for the single transition in a
mixed $XY$-Ising model which is 0.85.\cite{J.Lee91,Night95} It is
now believed that the XY character of the spins affects the Ising
chiral symmetry giving rise to those deviated critical exponents.
 Similarly, in the case of thin film studied here, we do not deal
with the discrete Ising model but rather an Ising-like Heisenberg
model. The Ising character due to chiral symmetry of the
transition at the surface is believed to be perturbed by the
continuous nature of Heisenberg spins. The transition of interior
layer shown in Fig. \ref{fig:gamma}(b) suffers similar but not the
same effects because of the absence of chiral symmetry on this
layer. So the value $\gamma/\nu$ is a little different. In the non
frustrated case shown in Fig. \ref{fig:gamma}(c), the deviation
from the 2D Ising universality class is less important because of
the absence of the chiral symmetry. This small deviation is
believed to stem mainly from the continuous nature of Heisenberg
spins.

To conclude this paragraph, we believe, from physical arguments
given above, that the deviation from 2D Ising universality class
of the transitions observed here is due to, in an decreasing order
of importance, the effect of the coupling between the continuous
degree of freedom of Heisenberg spin to the chiral symmetry, the
small Ising-like anisotropy and the film thickness.

\section{Concluding remarks}
We have studied, by means of a Green function method and MC
simulations, the Heisenberg spin model with an Ising-like
interaction anisotropy in thin films of stacked triangular
lattices. The two surfaces of the film are frustrated. We found
that surface spin configuration is non collinear when surface
antiferromagnetic interaction is smaller than a critical value
$J_s^c$. In the non collinear regime, the surface layer is
disordered at a temperature lower than that for interior layers
("soft" surface). This can explain the so-called "magnetically
dead surface" observed in some
materials.\cite{zangwill,bland-heinrich} The surface transition
disappears for $J_s$  larger than the critical value $J_s^c$.  A
phase diagram is established in the space ($T,J_s$). A good
agreement between the Green function method and the MC simulation
is observed. This is due to the fact that at high temperatures
where the transition takes place, the quantum nature of spins used
in the GF is lost so that we should find results of classical
spins used in MC simulations. We have also studied by MC histogram
technique the critical behavior of the phase transition using the
finite-size effects. The result of the ratio of critical exponents
$\gamma/\nu$ shows that the nature of the transition is
complicated due to the influence of several physical mechanisms.
The symmetry of the ground state alone cannot explain such a
result. We have outlined a number of the most relevant mechanisms.
Finally, we note that in surface magnetism the low surface
magnetization experimentally
observed\cite{zangwill,bland-heinrich} has been generally
attributed to the effects of the reduction  of magnetic moments of
surface atoms and/or the surface-localized low-lying magnon modes.
The model considered in this paper adds another origin for the low
surface magnetization: surface frustration.  It completes the list
of possible explanations for experimental observations in thin
films.

One of us (NVT) thanks 'World Laboratory' for financial support.
This work is supported by a contract between CNRS (France) and
VAST (Vietnam).

{}

\end{document}